%% file: AdvaniKitagawaSloczynski2019_EMCS.tex
\documentclass[12pt]{article}
\usepackage{AKS2018_style}
\usepackage{xr}
\usepackage{xcolor}

\begin{document}

\title{\textsc{\Large{Mostly Harmless Simulations?\\Using Monte Carlo Studies for Estimator Selection}}\thanks{This version: April 4, 2019. Previously circulated as `Mostly Harmless Simulations? On the Internal Validity of Empirical Monte Carlo Studies'. For helpful comments, we thank Thierry Magnac (Co-Editor), four anonymous referees, Alberto Abadie, Cathy Balfe, Richard Blundell, Colin Cameron, M{\'o}nica Costa Dias, Gil Epstein, Alfonso Flores-Lagunes, Ira Gang, Martin Huber, Justin McCrary, Blaise Melly, Mateusz My\'{s}liwski, Pedro Sant'Anna, Tony Strittmatter, Tim Vogelsang, Ed Vytlacil, Jeff Wooldridge, Tiemen Woutersen, and numerous seminar and conference participants. We also thank Michael Lechner and Blaise Melly for providing us with their codes, as well as Steven Karel and Francesco Pontiggia for assistance with the Brandeis HPC cluster. This research was supported by a grant from the CERGE-EI Foundation under a program of the Global Development Network (Grant No:~RRC12+09). All opinions expressed are those of the authors and have not been endorsed by CERGE-EI or the GDN\@. Advani also acknowledges support from Programme Evaluation for Policy Analysis, a node of the National Centre for Research Methods, supported by the ESRC (Grant No:~RES-576-25-0042). Kitagawa also acknowledges support from the ESRC through the ESRC Centre for Microdata Methods and Practice (cemmap) (Grant No:~RES-589-28-0001) and from the ERC through an ERC starting grant (Grant No:~EPP-715940). S\l{}oczy\'{n}ski also acknowledges support from the Foundation for Polish Science (FNP) through a START scholarship and from the Theodore and Jane Norman Fund. No authors are aware of any conflict of interest.}
}
\author{%
	\textsc{\large{Arun Advani,}}%
		\thanks{University of Warwick, CAGE, and Institute for Fiscal Studies.}
	~~\textsc{\large{Toru Kitagawa,}}%
		\thanks{University College London and cemmap.}
	~~\textsc{\large{and Tymon S\l{}oczy\'{n}ski}}%
		\thanks{Brandeis University and IZA\@. Correspondence:~Department of Economics \& International Business School, Brandeis University, MS 021, 415 South Street, Waltham, MA 02453. E-mail:~tslocz@brandeis.edu.}
}
\date{}

\begin{titlepage}
\maketitle

\renewcommand{\abstractname}{Summary}
\begin{abstract}
	\begin{normalsize}
	\input{abstract}

	\end{normalsize}
\end{abstract}
\thispagestyle{empty}

\end{titlepage}

\onehalfspacing
\setcounter{page}{2}

\section{Introduction}
\label{sec:intro}
\input{intro}

\section{EMCS Designs}
\label{sec:designs}
\input{designs}

\section{Theory}
\label{sec:theory}
\input{theory}

\section{Application}
\label{sec:application}
\input{application}

\section{Results}
\label{sec:results}
\input{results}

\section{Discussion}
\label{sec:discussion}
\input{discussion}

\pagebreak{}
\bibliographystyle{ECA_AA}
\bibliography{./EMCS_20190404}
\pagebreak{}

\appendix

\section*{Appendix}

\section{Theory} 
\label{sec:appendix_theory}
\input{appendix_theory}

\pagebreak
\section{Empirical Application: Estimators} 
\label{sec:appendix_estimators}
\input{appendix_estimators}

\pagebreak
\section{Empirical Application: Structured EMCS Procedure} 
\label{sec:appendix_strEMCSprocedure}
\input{appendix_strEMCSprocedure}

\end{document}

%% file: abstract.tex
We consider two recent suggestions for how to perform an empirically motivated Monte Carlo study to help select a treatment effect estimator under unconfoundedness.
We show theoretically that neither is likely to be informative except under restrictive conditions that are unlikely to be satisfied in many contexts.
To test empirical relevance, we also apply the approaches to a real-world setting where estimator performance is known.
Both approaches are worse than random at selecting estimators which minimise absolute bias.
They are better when selecting estimators that minimise mean squared error. 
However, using a simple bootstrap is at least as good and often better.
For now researchers would be best advised to use a range of estimators and compare estimates for robustness.\vspace{4pt}

\textit{JEL Classification:} C15, C21, C25, C52 \vspace{4pt}

\textit{Keywords:} empirical Monte Carlo studies, program evaluation, selection on observables, treatment effects

%% file: intro.tex
A large literature focuses on estimating average treatment effects under unconfoundedness (see, \textit{e.g.}, \citealt{BlundellCostaDias}, \citealt{ImbensWooldridge}, \citealt{AC2018}).\footnote{The unconfoundedness assumption may also be referred to as exogeneity, ignorability, or selection on observables.} Many estimators are available to researchers in this context, and many of these estimators have similar asymptotic properties. This can make it difficult to select which estimator to use. Monte Carlo studies are a useful tool for examining the small-sample properties of these estimation methods, which can guide estimator choice.\footnote{\label{fn:mcs}See, for example, \cite{Frolich2004,LuncefordDavidian2004,Zhao2004,Zhao2008,BussoDiNardoMcCrary_FiniteSample,Millimet,Austin2010,AbadieImbens2011,KPST,DiamondSekhon,HLW,BussoDiNardoMcCrary_PSMandReweighting,FHW2015}, and \cite{BCHL2018}, all studying the finite-sample performance of estimators of average treatment effects under unconfoundedness.} Early contributions, such as \cite{Frolich2004}, demonstrate estimator performance in stylised data generating processes (DGPs) which do not resemble any empirical settings. This reliance on unrealistic DGPs is criticised by \cite{HLW} and \cite{BussoDiNardoMcCrary_PSMandReweighting}. Both recommend that Monte Carlo studies should intend to replicate actual datasets of interest, although they suggest different procedures for doing this. 
\cite{HLW} describe this approach to examining the small-sample properties of estimators as an `empirical Monte Carlo study' (EMCS)\@. An important question is whether either type of EMCS can help applied researchers in choosing what estimator(s) to prefer in a given context. \cite{BussoDiNardoMcCrary_PSMandReweighting} indicate this might be possible, noting that their results `suggest the wisdom of conducting a small-scale simulation study tailored to the features of the data at hand'. Similarly, \cite{HLW} suggest that `the advantage [of an EMCS] is that it is valid in at least one relevant environment', \textit{i.e.}~that it is informative at least about the performance of estimators in the dataset on which it was conducted. 

In this paper we evaluate the premise that EMCS can be informative about the performance of estimators in the particular data which are the basis for the EMCS\@.
We first show theoretically that these approaches are expected to be informative only under very restrictive conditions. These conditions are unlikely to hold in many practical examples faced by a researcher. We then test EMCS performance in a real-world case where we know the actual behaviour of estimators. We find that in terms of selecting estimators on absolute bias they are often worse than choosing randomly. On mean squared error (MSE) they perform better than random, but no better than selecting an estimator based on simple bootstrap estimates of MSEs. Their performance in absolute terms may also still be poor. 

The first type of EMCS we consider is the \emph{placebo} EMCS \citep{HLW}.\footnote{It is also applied by \cite{LechnerWunsch,HLM2016,FHW2015,LS2017}, and \cite{BCHL2018}. A related approach is proposed by \cite{SynthVal}.} This proposes a way to assign `realistic placebo treatments among the non-treated', using information about the predictors of treatment status in the original data. It then tests how well estimators can recover the zero effect of the placebo treatment. The performance of estimators in this exercise is hypothesised to be informative about their performance in the original data.

The second type we describe as the \emph{structured} EMCS\@. An exercise of this type is undertaken by \cite{BussoDiNardoMcCrary_PSMandReweighting}.\footnote{A similar approach is also used by \cite{AbadieImbens2011}, \cite{Lee2013}, and \cite{DRR2015}.}
Here a parameterised approximation of the original data generating process is created, using functional form assumptions about the distributions of observed covariates. Parameters of their marginal (or conditional) distributions are estimated from the original data. Samples can be drawn from this approximate DGP, to which the estimators can then be applied. Since the treatment effect in this DGP can be calculated directly from knowledge of the parameters, performance of the estimators in these samples can be measured. The performance of estimators in this exercise is also hypothesised to be informative about their performance in the original data.

To examine whether or not EMCS can correctly choose a best performing estimator, for various definitions of performance, we first focus on a simple example with two estimators that have Gaussian sampling distributions.
We show analytically that both these approaches will only be guaranteed to correctly select the preferred estimator if they can correctly reproduce both the biases and the ordering of the variances of estimators. These are restrictive conditions that we show can easily fail in practical applications, such as when the EMCS procedures fail to recover heteroskedastic errors or misspecify the regression equations or propensity scores. 
In two sets of simulations based on a stylised DGP, both approaches select the better estimator less than 3\% of the time, much worse than 50\% achievable by selecting randomly.

To study the extent of the problem in a real-world circumstance, we apply both methods to the National Supported Work (NSW) Demonstration data on men, previously analysed by \cite{LaLonde1986}, \cite{HeckmanHotz1989}, \cite{DehejiaWahba1999,DehejiaWahba2002}, \cite{SmithTodd2001,SmithTodd2005}, and many others. In these data participation in a job training programme was randomly assigned, so the treatment effect of the programme can be estimated by comparing sample means. \cite{LaLonde1986} used these data to test the performance of estimators at reproducing this treatment effect when an artificial comparison group (rather than the experimental controls) was used. We instead use the data to test how well the two EMCS procedures can inform us about the performance of the estimators: Can EMCS tell us which estimator to use? On average how much worse than the optimal estimator is the one chosen by EMCS\@? How well can EMCS reproduce the ranking of performance across all estimators? 

Applying the two EMCS procedures we find three main results. First, in terms of absolute bias, the EMCS procedures are no better, and often noticeably worse, than selecting an estimator at random. In two out of three cases we study, the rankings produced are \emph{negatively} correlated with the true ranking. In one case the preferred estimator selected by EMCS is on average 30--37 times worse than the actual best estimator.

Second, EMCS does better at reproducing the performance of estimators in terms of MSE\@. This is because the MSEs of the estimators are mostly driven by their variances, and EMCS appears more effective at capturing variances. The rankings of estimators are consistently positively correlated with the true rankings, although the estimator preferred by EMCS has an MSE up to twice as high as the best estimator.

Third, given the variance result, we also compare EMCS procedures to choosing estimators based on performance criteria estimated from a simple nonparametric bootstrap. We find that the bootstrap is as good, and often much better, than either of the EMCS procedures. Hence even when the procedures are somewhat informative, they are not superior to a procedure that relies on fewer design choices.

These results are unfortunate, but nevertheless important. They caution against treating either of these approaches as general solutions to the problem of estimator choice. There remains no silver bullet that can assist empirical researchers with the `right' or `best' estimator for a particular context. In the absence of a clear choice driven by research design, the best advice at this stage is likely to be implementing a number of estimators, and then considering the range of estimates provided, as \cite{BussoDiNardoMcCrary_PSMandReweighting} also suggest.

Our results also have implications for researchers studying the small-sample properties of treatment effect estimators (see footnote \ref{fn:mcs}). It has been argued that `it is preferable to study DGPs that are empirically relevant' (\citealp{BussoDiNardoMcCrary_PSMandReweighting}). Our theoretical and empirical results suggest there is little support for this claim. We show theoretically that misspecification in the construction of the DGP can lead the ranking of estimators to be incorrect for the original dataset. In our empirical example, we see that EMCS is not better than using a bootstrap (and sometimes not better than random) to predict performance in the data on which the EMCS was performed. There seems to be little reason to then think it is particularly informative about performance in other unrelated real datasets, \emph{i.e.}~that testing small-sample properties of estimators in `real data' is necessarily better than in completely artificial data. A more fruitful path might be to test sensitivity of estimator performance to parameters of the simulation, such as sample size and the degree of heteroskedasticity. This approach is also taken by \cite{HLW}, and might be more helpful in understanding what characteristics of samples most affect the performance of particular estimators.

%% file: designs.tex
We first describe the two main approaches to conducting an EMCS, namely the placebo design of \cite{HLW} and the structured design of \cite{BussoDiNardoMcCrary_PSMandReweighting}. 
In either EMCS design, one simulates many `empirical Monte Carlo' replication samples from a known data generating process. By implementing the estimators on the simulated replications, one obtains estimates of the sampling distributions and performance criteria (\emph{e.g.},~MSEs) of the estimators, according to which one ranks the candidate estimators. Note that the researcher needs to make a choice of what criteria to use to rank estimators.

\subsection{The Placebo Design}
\label{subsec:design_pla}

The idea of the placebo design is to assign placebo treatments to some control observations, so that by construction the treatment effect is zero, and to then attempt to recover this effect.\footnote{A similar approach is developed by \cite{BertrandEtAl2004} who study inference in difference-in-differences methods using simulations with randomly generated `placebo laws' in state-level data, \textit{i.e.}~policy changes which never actually happened. For follow-up studies, see \cite{Hansen2007}, \cite{CameronEtAl2008}, and \cite{BCJ2018}.} 
In particular, covariates and outcomes $(\mathbf{X}_{i},Y_{i})$ are first drawn jointly by sampling (with replacement) from the empirical distribution of control observations.
Using the original dataset, the propensity score is estimated (\textit{e.g.}, using a logit model). The estimated parameters of this model $\boldsymbol{\hat{\phi}}$ are then used to assign placebo treatments to the generated sample in the following way:
\begin{eqnarray}
	D_{i}    & = & 1[ S_{i}>0 ], \\
	S_{i} & = & \pi+\lambda\mathbf{X}_{i}\boldsymbol{\hat{\phi}}+\epsilon_{i}, \label{eq:latent}
\end{eqnarray}
where $\epsilon_{i}$ is an iid error, and both $\pi$ and $\lambda$ are additional parameters to be selected. While $\pi$ shifts the proportion of observations that are treated, $\lambda$ controls the extent of selection: with $\lambda=1$ selection on observables takes the same form in the Monte Carlo samples as in the original dataset.

\subsection{The Structured Design}
\label{subsec:design_str}

The idea of the structured design is instead to create a parameterised approximation to the original (unknown) data generating process, and then draw samples from the approximated process. To begin, a fixed number of treated and control observations are created, to match the number of each in the original dataset. Covariates and outcome variables are then drawn from parameterised distributions where the parameters are estimated from the original dataset. For example, the variable \texttt{black} might come from a Bernoulli with mean estimated from the data, and the variable \texttt{earnings} from a log-normal distribution with mean and variance estimated from the data. The parameters of these distributions are typically estimated conditional on treatment status. Parameters of some distributions might also be conditional on the value of other variables; \textit{e.g.}, earnings might be conditional on race as well as treatment status. More conditioning will improve the match of the joint distribution of simulated data to the joint distribution of the original data, but will increase the number of parameters that need to be estimated.

%% file: theory.tex
To understand the conditions under which an EMCS might be informative about the preferred estimator in some particular dataset, we first construct a simple example. Here we have only two estimators, with a straightforward and restricted joint sampling distribution (bivariate Gaussian). This bivariate Gaussian setting mimics an ideal situation in which the finite sample distribution of the estimators is well approximated by their asymptotic distribution. We show that even in such an ideal, large sample situation, EMCS can fail to select the best estimator if the bias in any one of the estimators or the ranking of variances is not correctly replicated in the simulated samples.
We provide simple common cases for treatment effect estimation in which failure to capture the biases and heteroskedasticity contaminates EMCS, and provide results from a simple simulation illustrating this. We then extend the example to the case of more than two estimators.

\subsection{Simple Example: Two Estimator Case}
\label{subsec:SimpleExample-TwoEstimatorCase}

Suppose the researcher wants to rank two estimators $\hat{\theta}_1$ and $ \hat{\theta}_2 $ according to their statistical performances under repeated sampling. These estimators are estimating the same object of interest $\theta \in \mathbb{R}$, but their constructions are different.  For simplicity of the illustration, assume that the joint sampling distribution of the two estimators is bivariate Gaussian:
\begin{equation}
\begin{pmatrix} \hat{\theta}_1 \\ \hat{\theta}_2 \end{pmatrix} \sim \mathcal{N} \left( \begin{pmatrix} \theta_1 \\ \theta_2 \end{pmatrix},  \Sigma_{n}  \right), \label{eq:normal1}
\end{equation}
where $\Sigma_n = n^{-1} \Sigma$, $\Sigma= \begin{pmatrix} \sigma_1^2 & \sigma_{12} \\ \sigma_{12} & \sigma_2^2 \end{pmatrix}$, and $n$ is the sample size. Here, our implicit assumption is that the estimators $(\hat{\theta}_1, \hat{\theta}_2)$ converge to $(\theta_1, \theta_2)$ at $\sqrt{n}$-rate. Let $\theta^0$ be the true value of the parameter of interest. We allow $\hat{\theta}_1$ and/or $\hat{\theta}_2$ to be biased so that $\theta_1$ and/or $\theta_2$ can differ from $\theta^0$.

We rank these estimators according to their statistical performances. Given that we often assess the performance of an estimator by its mean squared error (MSE) or mean absolute error (MAE), we may, for instance, rank the estimators according to their MSEs or MAEs.\footnote{\label{fn:poc}MSE and MAE criteria do not take into account the dependence of the estimators. One way to rank the estimators that takes into account their dependence is based on \textit{the probability of being closer to the truth}, $\Pr(|\hat{e}_1| \leq |\hat{e}_2| )$, where $\hat{e}_1 = \hat{\theta}_1 - \theta^0$ and $\hat{e}_2 = \hat{\theta}_2 - \theta^0$ are the estimation errors of the two estimators. That is, $\hat{\theta}_1$ is preferred to $\hat{\theta}_2$ if $\Pr\left(|\hat{e}_1| \leq |\hat{e}_2| \right) > 1/2$ and $\hat{\theta}_2$ is preferred to $\hat{\theta}_1$ if $\Pr\left(|\hat{e}_1| \leq |\hat{e}_2| \right) < 1/2$. Considering this criterion instead of MSE does not affect the main results in our simple example. } Given the Gaussian assumption, the MSE of each estimator, $j=1,2$, is
\[
MSE(\hat{\theta}_j) = (\theta_j - \theta^0)^2 + n^{-1} \sigma_j^2.
\]

We denote by $j_0 \in \{ 1,2 \} $ the index of the strictly preferred estimator, assuming it exists. Ranking the estimators is difficult in practice since we do not know the mean and variances of the estimators as well as the true value of $\theta$. Proposals of the EMCS literature aim to infer a best performing estimator $j_0$ by estimating the sampling distribution of $\hat{\theta}_1$ and $\hat{\theta}_2$ via some Monte Carlo studies.  For simplicity, we assume that the estimators simulated in EMCS also follow bivariate Gaussian,
\begin{equation}
\begin{pmatrix} \hat{\theta}_1^{\ast} \\ \hat{\theta}_2^{\ast} \end{pmatrix} \sim \mathcal{N} \left( \begin{pmatrix} \tilde{\theta}_1 \\ \tilde{\theta}_2 \end{pmatrix},  \tilde{\Sigma}_n  \right), \label{eq:normal2}
\end{equation}
where $\tilde{\Sigma}_n = a_n^{-1} \tilde{\Sigma}$, $\tilde{\Sigma}=\begin{pmatrix} \tilde{\sigma}_1^2 & \tilde{\sigma}_{12} \\ \tilde{\sigma}_{12} & \tilde{\sigma}_2^2 \end{pmatrix}$, and $a_n$ is the size of a simulated sample that may differ from the size of the original sample $n$. The underlying parameters in EMCS, $(\tilde{\theta}_1, \tilde{\theta}_2, \tilde{\Sigma})$, generally depend on the original sample, but we assume for simplicity that the dependence is negligible and they can be treated as constants. EMCS computes $\hat{\theta}_1^{\ast}$ and $\hat{\theta}_2^{\ast}$ repeatedly using simulated samples of size $a_n$ drawn from a data generating process with the parameter value set at known value $\tilde{\theta}^{0}$. For instance, the placebo EMCS approach of \cite{HLW} sets $\tilde{\theta}^{0} = 0$ and $a_n \leq n_0$, the size of the control group in the original data. The approach of structured EMCS sets $\tilde{\theta}^0$ at an estimate of $\theta^0$ constructed from the original sample. In implementing EMCS, we do not have to know the mean and variance parameters of $(\hat{\theta}_1^{\ast}, \hat{\theta}_2^{\ast})$, and they can be estimated with arbitrary accuracy based on the simulated estimators. EMCS accordingly obtains the MSE of each estimator, $j=1,2$, by
\[
\widehat{MSE}(\hat{\theta}_j) = (\tilde{\theta}_j - \tilde{\theta}^0 )^2 + a_n^{-1} \tilde{\sigma}_j^2.
\]

We denote by $\hat{j}_0$ the index for a best performing estimator estimated from EMCS, $\hat{j}_0 \equiv \arg \min_j \widehat{MSE}(\hat{\theta}_j)$. To assess the validity of EMCS, we define a criterion of \textit{EMCS-validity} by the probability that $\hat{j}_0$ coincides with $j_0$, $\Pr(\hat{j}_0 = j_0)$, where the probability is evaluated under repeated sampling of the original samples. In the examples to follow, we investigate how this criterion of EMCS-validity becomes one or zero depending on the parameter values in the bivariate Gaussian distributions of (\ref{eq:normal1}) and (\ref{eq:normal2}). We assume away the dependence of the parameters in (\ref{eq:normal2}) on the original sample for simplicity of illustration. In such a case the MSE estimates in EMCS and resulting selection of a best estimator $\hat{j}_0$ are nonrandom when the number of Monte Carlo iterations is large enough. The criterion of EMCS-validity in this case is either 1 or 0.

We can also consider the \textit{average regret} type criterion such as $\E(MSE(\hat{\theta}_{\hat{j}_0}) - MSE(\hat{\theta}_{j_0})) \geq 0$
to quantify EMCS-validity. Here, the expectation concerns the sampling distribution of EMCS's selection of an optimal estimator $\hat{j}_0$. This average regret criterion can quantify severity of a wrong choice of the estimators in terms of how much MSE is on average sacrificed relative to the true best-performing estimator.

\subsubsection{Scenario 1}
Denote the biases in $(\hat{\theta}_1, \hat{\theta}_2)'$ by $\textbf{b}=(b_1,b_2)' = (\theta_1 - \theta^0, \theta_2-\theta^0)'$ and the biases in $(\hat{\theta}_1^{\ast}, \hat{\theta}_2^{\ast})'$ by $\tilde{\textbf{b}} = (\tilde{b}_1, \tilde{b}_2)' =  (\tilde{\theta}_1 - \tilde{\theta}^0, \tilde{\theta}_2-\tilde{\theta}^0)'$. We start with a scenario in which $(\hat{\theta}_1, \hat{\theta}_2)$ are unbiased and the distribution of $(\hat{\theta}_1^{\ast}, \hat{\theta}_2^{\ast})$ well replicates the distribution of $(\hat{\theta}_1, \hat{\theta}_2)$ in the following sense:
\begin{equation}
\textbf{b} = \tilde{\textbf{b}} = \mathbf{0}, \mspace{30mu} \Sigma = \tilde{\Sigma}. \label{eq:case1}
\end{equation}
Here, the biases and the sample-size-adjusted variances of the estimators simulated in EMCS coincide with those of the estimators in the original data generating process. Note that the true parameter value assumed in EMCS, $\tilde{\theta}^0$, does not have to agree with the true parameter value in the original sampling process, $\theta^0$.

In the current scenario, the ranking of the true MSEs clearly coincides with the ranking of the MSE estimates in EMCS, implying $\Pr(\hat{j}_0 = j_0)=1$. This is a benchmark case in which EMCS works. The next two scenarios show that once we depart from the assumptions in (\ref{eq:case1}), EMCS can be no longer valid.

\subsubsection{Scenario 2}
Assume that the estimators are free from biases both in the original data generating process and EMCS, $\textbf{b} = \tilde{\textbf{b}} = \textbf{0}$, but EMCS fails to replicate the normalised covariance matrix of the estimators, $\Sigma \neq \tilde{\Sigma}$. In this case, the MSE estimates in EMCS correctly rank the true MSEs of the estimators (assuming $\sigma_1^2 \neq \sigma_2^2$) if and only if the ordering of the variances of the two estimators agrees between the original sampling process and the simulated sampling process, \textit{i.e.}~$(\sigma_1^2 - \sigma_2^2)(\tilde{\sigma}_1^2 - \tilde{\sigma}_2^2)>0$. Otherwise, EMCS  reverses the ranking of the estimators and incorrectly selects a suboptimal estimator as optimal, $\Pr(\hat{j}_0 = j_0)=0$.

Hence, even when EMCS well replicates the biases of the estimators, it can fail to select a best performing estimator due to an incorrect variance ordering.

\subsubsection{Scenario 3}
In the third scenario, we assume that EMCS correctly replicates the variance ordering of the estimators, \textit{i.e.}~$(\sigma_1^2 - \sigma_2^2)(\tilde{\sigma}_1^2 - \tilde{\sigma}_2^2) > 0$, but fails to replicate the biases, $(b_1,b_2) \neq (\tilde{b}_1, \tilde{b}_2)$. To be specific, we set $(\tilde{b}_1, \tilde{b}_2) = (0,0)$, but $(b_1,b_2) = (0, b_2), b_2 \neq 0$. This can correspond to a situation that the estimator 1 is correctly specified and has no bias, whereas estimator 2 is misspecified and is subject to bias in the original data generating process. EMCS, however, fails to capture the misspecification bias in estimator 2.

Suppose $\sigma_1^2 > \sigma_2^2$ holds. The true MSEs are $MSE(\hat{\theta}_1)=n^{-1} \sigma_1^2$ and $MSE(\hat{\theta}_2)= b_2^2 +n^{-1} \sigma_2^2$,
while the MSE estimates in EMCS are $\widehat{MSE}(\hat{\theta}_1) = a_n^{-1} \tilde{\sigma}_1^2$ and $\widehat{MSE}(\hat{\theta}_2) = a_n^{-1} \tilde{\sigma}_2^2$. Since we assumed that EMCS correctly replicates the variance of the estimators, EMCS selects $j=2$ as a best estimator. This selection of the estimator is indeed misleading if $b_2$ is far from zero, since if $|b_2| > \sqrt{\frac{\sigma_1^2 - \sigma_2^2}{n}}$, $\hat{\theta}_1$ outperforms $\hat{\theta}_2$ in terms of MSE.

This scenario highlights that EMCS-based selection of the estimator can fail if any one of the estimators is misspecified and the simulation design in EMCS does not replicate the misspecification bias.

\subsection{Are Scenarios 2 and 3 Relevant in Treatment Effect Estimation?}
\label{subsec:AreScenarios2And3RelevantInTreatmentEffectEstimation}
We next provide simple but empirically relevant examples where we focus on the estimation of treatment effects, and show that both types of EMCS may yield misleading choices of the estimators for the reasons illustrated in Scenarios 2 and 3 above.

Data are given by a random sample of $\{ (Y_i, D_i, X_i): i=1,\dots, n \}$, where $Y_i \in \mathbb{R}$ is unit $i$'s observed post-treatment outcome, $D_i \in \{ 0,1 \}$ is her treatment status, and $X_i \in \mathbb{R}^{d_x}$ is a vector of her pre-treatment characteristics whose support is assumed to be bounded. We denote unit $i$'s potential outcomes by $(Y_i(1),Y_i(0))$. We assume the unconfoundedness assumption, $(Y(1),Y(0)) \perp D | X$, throughout. The propensity score is denoted by $e(x) = \Pr (D=1 | X=x)$.

\subsubsection{An Example for Scenario 2}
\label{subsubsec:AnExampleForScenario2}
To keep our example as simple as possible, consider the following data generating processes:
\begin{align}
\E(Y(1)|X=x) &= \beta_0 + \beta_1 + x'\beta_2, \label{eq:DGP1} \\
\E(Y(0)|X=x) &= \beta_0  + x'\beta_2, \notag \\
Var(Y(1)|X=x)  & =  c \sigma_{\epsilon}^2, \mspace{20mu} Var(Y(0)|X=x)  =  \sigma_{\epsilon}^2, \mspace{20mu} c > 0, \notag \\
e(x) & = \gamma_0 + x' \gamma_1.  \notag
\end{align}
The specified mean equations for both potential outcomes imply that the conditional average treatment effects are homogeneous over observed characteristics and equal to $\beta_1$. The potential outcomes are heteroskedastic if $c \neq 1$. We assume a linear probability for the propensity score in order to simplify analytical comparisons of the variances of the estimators we introduce below.

Suppose that the parameter of interest is the population average treatment effect for the treated (ATT), $\theta^0 = \E(Y(1) - Y(0)|D=1)$. Since specification (\ref{eq:DGP1}) implies homogeneous conditional average treatment effects, $\E(Y(1) - Y(0)|X=x) = \beta_1$, the true value of ATT is $\theta^0 = \beta_1$.

We consider two different estimators to estimate the population ATT\@. The first estimator $\hat{\theta}_1$ is a semiparametric estimator for ATT, which is consistent without assuming functional forms for the outcome and propensity score equations, and asymptotically attains the semiparametric efficiency bound (SEB) of ATT derived by \cite{Hahn1998}. Estimators that attain this property include the inverse probability weighting (IPW) estimator with nonparametrically estimated propensity scores \citep{HIR2003}, doubly robust estimators of \cite{Hahn1998}, covariate or propensity score matching estimators with a single covariate \citep{AbadieImbens2006,AbadieImbens2016}, and covariate balancing estimators of \cite{ChenHongTarozzi2008} and \cite{GrahamEtAl2012,GrahamEtAl2016}. We can set any one of these estimators as our first estimator without affecting the analysis below.

We specify the second estimator $\hat{\theta}_2$ as the ordinary least squares estimator of $\beta_1$ in the following regression equation:
\begin{equation}
Y_i = \beta_0 + \beta_1 D_i + X_i'\beta_2 + \epsilon_i, \mspace{20mu} \E(\epsilon_i | D_i, X_i) =0. \label{eq:simpleOLS}
\end{equation}
In other words, $\hat{\theta}_2=\hat{\beta}_{1,OLS}$. The specification of (\ref{eq:DGP1}) implies that $\hat{\theta}_2$ is unbiased and consistent for the population ATT, $\theta^0$. We consider a situation in which the finite sample distribution of $(\hat{\theta}_1, \hat{\theta}_2)$ is well approximated by its large sample normal approximation, \textit{i.e.}
\begin{equation}
\begin{pmatrix}  \hat{\theta}_1 \\ \hat{\theta}_2 \end{pmatrix} \sim \mathcal{N} \left( \begin{pmatrix} \theta^0 \\ \theta^0 \end{pmatrix}, \frac{1}{n} \begin{pmatrix}  \sigma_1^2 & \sigma_{12} \\ \sigma_{12} & \sigma_2^2 \end{pmatrix} \right), \notag
\end{equation}
where $\sigma_1^2$ is the asymptotic variance of $\sqrt{n}(\hat{\theta}_1 - \theta^0)$ given by SEB for ATT without the knowledge of propensity scores, and $\sigma_2^2$ is the asymptotic variance of $\sqrt{n}(\hat{\theta}_2 - \theta^0)$. Under the current specification, they are obtained as
\begin{align}
\sigma_1^2 & = \frac{\sigma_{\epsilon}^2}{\Pr(D=1)} \left[ c + \E\left( \frac{e(X)}{1-e(X)} |  D=1 \right)\right], \label{eq:SEB1} \\
\sigma_2^2 & = \frac{\sigma_{\epsilon}^2}{\Pr(D=1)} \left[ c \cdot \frac{\E((1-e(X))^2 | D=1)}{[\E(1-e(X)|D=1)]^2} + \frac{\E(e(X)(1-e(X))|D=1)}{[\E(1-e(X)|D=1)]^2}\right]. \label{eq:sigma2}
\end{align}
See Appendix~\ref{sec:appendix_theory} for their derivations.

When $Y(1)$ and $Y(0)$ share the variance ($c=1$), it can be shown that the OLS estimator is more efficient than the semiparametric estimator, $\sigma_2^2 < \sigma_1^2$, due to exploitation of the correct functional form of the regression equation. In contrast, if the variance of the treated outcome is higher than the variance of the control outcome ($c >1$), the simple OLS estimator that does not take into account the heteroskedastic errors can become less efficient than the semiparametric estimator. Specifically, we show in Appendix~\ref{sec:appendix_theory} that
\begin{align}
\sigma_2^2 & > \sigma_1^2 \mspace{15mu} \text{iff $c > \frac{\Delta_1}{\Delta_2}+1$, where} \label{eq:condition_c} \\
&\Delta_1 = \E \left[ \frac{1}{1-e(X)} | D=1 \right] - \frac{1}{\E(1-e(X)|D=1)} \geq 0, \notag \\
& \Delta_2 = \frac{\E((1-e(X))^2 | D=1)}{[\E(1-e(X)|D=1)]^2} - 1 \geq 0. \notag
\end{align}
Hence, if the degree of heteroskedasticity satisfies the condition in (\ref{eq:condition_c}),
the semiparametric estimator $\hat{\theta}_1$ is strictly preferred to the OLS estimator $\hat{\theta}_2$.

Given that $c$ meets (\ref{eq:condition_c}), consider applying the placebo EMCS proposed in \cite{HLW}. We assume that the two estimators are centred at zero and their simulated distributions can be well approximated by bivariate Gaussian,
\begin{equation}
\begin{pmatrix}  \hat{\theta}_1^{\ast} \\ \hat{\theta}_2^{\ast} \end{pmatrix} \sim \mathcal{N} \left( \begin{pmatrix} 0 \\ 0 \end{pmatrix}, \frac{1}{n_0} \begin{pmatrix}  \tilde{\sigma}_1^2 & \tilde{\sigma}_{12} \\ \tilde{\sigma}_{12} & \tilde{\sigma}_2^2 \end{pmatrix} \right), \notag
\end{equation}
where $n_0$ is the sample size of control group in the original sample. Suppose also that the propensity scores used to generate the placebo treatment coincide with the true propensity scores in the original data. Since the placebo treated group is generated from the original control group, it fails to replicate the variance of the treatment outcomes in the original data. As a result, the variances of $\sqrt{n_0} \hat{\theta}_1^\ast$ and $\sqrt{n_0} \hat{\theta}_2^\ast$ are given by the homoskedastic version ($c=1$) of (\ref{eq:SEB1}) and (\ref{eq:sigma2}),
\begin{equation}
\tilde{\sigma}_1^2 = \frac{\sigma_{\epsilon}^2}{\widetilde{\Pr}(D=1)} \tilde{\E} \left[ \frac{1}{1-e(X)} | D=1 \right] \geq  \tilde{\sigma}_2^2  = \frac{\sigma_{\epsilon}^2}{\widetilde{\Pr}(D=1)} \cdot  \frac{1}{\tilde{\E}(1-e(X)|D=1)}, \label{eq:EMCSvariances}
\end{equation}
where $\widetilde{\Pr}$ and $\tilde{\E}$ are the probability and expectation with respect to the sampling distribution specified in the placebo EMCS\@. This inequality is strict if $e(X)|D=1$ is nondegenerate. EMCS therefore incorrectly selects the OLS estimator $\hat{\theta}_2$ as a preferred estimator.

The underlying mechanism for why EMCS goes wrong is in line with Scenario 2 in the previous subsection. Even in a rather ideal situation where EMCS well replicates the unbiasedness of the estimators, artificially creating a placebo treated group from the control group in the original sample distorts the variance ordering among the estimators.

Exactly the same reasoning can also invalidate structured EMCS designs if the estimated data generating process from which the data are to be simulated ignores or fails to replicate the underlying heteroskedasticity of the potential outcome distributions.

This problem can be seen in a simple simulation study. We draw 1,000 samples from a data generating process of the form given by equation~(\ref{eq:DGP1}) with 1,000 observations per sample.\footnote{See \cite{AKS2019_SA} for details: Appendix~\ref{sec:appendix_stylisedsimulations} for our procedure and parameter values; Appendix~\ref{sec:appendix_stylisedtables} for simulation results.} For each sample we run 1,000 replications of the placebo and structured EMCS procedures, considering IPW and OLS as our two estimators. This gives us `the true MSE' for each estimator (based on the original samples) as well as 1,000 estimates of the MSE for each combination of an estimator and an EMCS design. Looking at a simple count of how many times each procedure selects the right estimator, we see that the placebo approach selects the superior estimator only 19 times (1.9\% of the time) and the structured approach is little better at 30 times (3.0\%). This compares with 97.6\% and 100\% for the placebo and structured procedures, respectively, when there is no heteroskedasticity. Of course this is a single example, and in a very stylised context; in Section~\ref{sec:application} we will see that the performance of these methods is also poor in a `real-world' example.

\subsubsection{An Example for Scenario 3}
We shift our focus to Scenario 3. We now introduce a bias in one of the estimators in the original data generating process. For this purpose, we maintain the two estimators as in the previous example, but alter the potential outcome equations from (\ref{eq:DGP1}) with
\begin{align}
\E(Y(1)|X=x) &= \beta_0 + \beta_1 + x' \beta_t, \label{eq:DGP2} \\
\E(Y(0)|X=x) &= \beta_0 + x'\beta_c, \notag
\end{align}
with distinct slopes, $\beta_t \neq \beta_c$. This causes the regression specification of (\ref{eq:simpleOLS}) to be misspecified so that $\hat{\theta}_2$ is no longer consistent for the population ATT, $plim_{n \to \infty} \hat{\theta}_2 = \theta_2 \neq \theta^0 = \beta_1 + \E(X'|D=1)(\beta_t - \beta_c)$. See, \textit{e.g.}, \cite{Sloczynski2018} for analytical characterizations of the bias. On the other hand, the semiparametric estimator $\hat{\theta}_1$ remains consistent and semiparametrically efficient (asymptotically attains SEB)\@.
Hence, assuming that the finite sample distribution of $(\hat{\theta}_1, \hat{\theta}_2)$ is well approximated by its asymptotic normal approximation, we have
\begin{equation}
\begin{pmatrix}  \hat{\theta}_1 \\ \hat{\theta}_2 \end{pmatrix} \sim \mathcal{N} \left( \begin{pmatrix} \theta^0 \\ \theta^0 + b_2 \end{pmatrix}, \frac{1}{n} \Sigma \right). \notag
\end{equation}
As we argued in Scenario 3 above, the bias in $\hat{\theta}_2$ makes $\hat{\theta}_2$ inferior to unbiased estimator $\hat{\theta}_1$  even when $\sigma_2^2 < \sigma_1^2$ if $b_2$ or the sample size is sufficiently large.

In the placebo EMCS procedure of \cite{HLW}, the fact that the placebo treated group is generated from the original control group removes the misspecification issue of the OLS estimator caused by the non-parallel potential outcome equations. Hence, $\hat{\theta}_2^{\ast}$ behaves as a correctly specified OLS estimator with homoskedastic errors, and the simulated distribution of $\hat{\theta}_2^{\ast}$ fails to replicate the bias in $\hat{\theta}_2$. Since the variance ordering in EMCS obtained in (\ref{eq:EMCSvariances}) is preserved in the current example, EMCS erroneously concludes that the OLS estimator $\hat{\theta}_2$ dominates the semiparametric estimator $\hat{\theta}_1$.

In case of structured EMCS procedures, if the data generating process from which Monte Carlo samples are drawn is estimated under misspecification, the structured EMCS misleads the estimator selection for exactly the same reason. For example, if one were to construct the Monte Carlo data generating process using linear regressions additive in $D_i$, structured EMCS will then wrongly conclude that the OLS estimator $\hat{\theta}_2$ outperforms the semiparametric estimator $\hat{\theta}_1$.

Again we perform a simple simulation, analogous to the previous subsection but modifying the potential outcome equations as given by equation~(\ref{eq:DGP2}). We perform 1,000 replications of each EMCS procedure using the same estimators, and then compare in how many cases the EMCS correctly selects the estimator with the lower MSE. Again the performance of EMCS is rather poor: placebo EMCS correctly selects IPW 2.3\% of the time, and structured EMCS is correct only .2\% of the time. See \cite{AKS2019_SA} for further details.

\subsection{More Than Two Estimators}
\label{subsec:MoreThanTwoEstimators}

Applications of EMCS often consider comparing more than two estimators. Fragility of EMCS-based estimator selection highlighted in the two estimator examples above naturally carries over to settings with more than two estimators, since ranking over multiple estimators consists of transitive pairwise rankings of any two candidate estimators.

The Monte Carlo exercises and the empirical application below consider a setting with seven estimators in the context of program evaluation with observational data. Let $(\hat{\theta}_1, \dots, \hat{\theta}_J)$ be the pool of $J$ candidate estimators, and let the purpose of EMCS be to obtain a \textit{complete} ordering among these $J$ estimators according to the MSE criteria.

The EMCS-validity criteria introduced above, $\Pr(\hat{j}_0 = j_0)$ and $\E(MSE(\hat{\theta}_{\hat{j}_0}) -$ $MSE(\hat{\theta}_{j_0}))$, can be straightforwardly extended to the case with several estimators. In addition, to measure similarity or dissimilarity between the true ranking and estimated rankings in EMCS, it can be of interest to look at the distribution of the Kendall's tau,
\begin{equation}
\hat{\tau} = \frac{2}{J(J-1)} \sum_{i < j} 1\{ (\rho(i) - \rho(j))(\hat{\rho}(i) - \hat{\rho}(j)) > 0 \} \notag
\end{equation}
where $\rho(j)$ and $\hat{\rho}(j)$, $j \in \{1, \dots, J\}$, are the ranks of estimator $j$ with respect to the true MSE and estimated MSE in EMCS, respectively. Noting $\hat{\tau} \in [-1,1]$ has a distribution under repeated sampling, its mean or other location parameters can summarise how well EMCS can assess the relative performances among the candidate estimators.

%% file: application.tex
To demonstrate the empirical relevance of the theoretical results discussed above, and consider the extent to which they might be a problem in practice, we provide an application of EMCS procedures to a real-world dataset. In these data we have an experimental estimate of the treatment effect. 
By (initially) treating the experimental estimate as the true treatment effect, the aim is to show whether (or not) EMCS procedures can accurately recover the ranking of estimators that we see from the experiment. We first discuss the data used, then our approach, next the estimators, and finally the details of how the EMCS procedures were conducted.

\subsection{Data and Context}
\label{subsec:data}

We focus on the data for men from \cite{LaLonde1986}, used also by \cite{HeckmanHotz1989}, \cite{DehejiaWahba1999,DehejiaWahba2002}, and \cite{SmithTodd2001,SmithTodd2005}.\footnote{Recent work by \cite{CalonicoSmith2017} highlights the effects of the NSW programme for women. Prior to this women were largely ignored in the NSW literature subsequent to \cite{LaLonde1986} because the analysis datafile for women was not preserved.} A subset of these data comes from the National Supported Work (NSW) Demonstration, which was a work experience programme that operated in the mid-1970s at 15 locations in the United States (for a detailed description of the programme see \citealp{SmithTodd2005}). This programme served several groups of disadvantaged workers, such as women with dependent children receiving welfare, former drug addicts, ex-convicts, and school drop-outs. Unlike many social programmes, the NSW implemented random assignment among eligible participants. This random selection allowed for straightforward evaluation of the programme via a comparison of mean outcomes in the treatment and control groups.

In an influential paper, \cite{LaLonde1986} uses the design of this programme to assess the performance of a large number of nonexperimental estimators of average treatment effects, many of which are based on the assumption of unconfoundedness. He sets aside the original control group from the NSW data and creates several alternative comparison groups using data from the Current Population Survey (CPS) and the Panel Study of Income Dynamics (PSID), two standard datasets on the U.S. population. His key insight is that a `good' estimator should be able to closely replicate the experimental estimate of the effect of NSW using nonexperimental data. He finds that very few of the estimates are close to this benchmark. This result motivated a large number of replications and follow-ups, and established a testbed for estimators of average treatment effects under unconfoundedness (see, \textit{e.g.}, \citealt{HeckmanHotz1989}; \citealt{DehejiaWahba1999,DehejiaWahba2002}; \citealt{SmithTodd2001,SmithTodd2005}; \citealt{AbadieImbens2011}; \citealt{DiamondSekhon}). Like many other papers, we use the largest of the six nonexperimental comparison groups constructed by \cite{LaLonde1986}, which he refers to as CPS-1.

\subsection{Approach}
\label{subsec:approach}

In this paper we take the key insight of \cite{LaLonde1986} one step further. We treat the NSW--CPS data from \cite{LaLonde1986} as a finite population, with 185 treated observations and 7,660 comparison observations in our main example. This comes from taking the treated sample used by \cite{DehejiaWahba1999} and a trimmed version of the \mbox{CPS-1} dataset, where the literature suggests conditional independence might hold using the available conditioning variables.\footnote{We use a logit model to predict propensity to be in the experimental data (either as treatment or control) versus being in the CPS-1 data. We then drop all CPS-1 observations with propensity scores below the minimum or above the maximum in the experimental data. This is the trimmed CPS-1 dataset, which we then combine with the NSW treated observations from \cite{DehejiaWahba1999}.} From this we draw 1,000 samples, each composed of 100 treated observations and 1,900 comparison observations. We then implement the estimators described below. For each sample and each estimator we compute the difference between the estimate and the `true effect' (\$1,794), which comes from the experimental estimate of the impact of NSW on earnings. With 1,000 such differences for each estimator, we can compute the MSE and other performance measures for that estimator in these data. Then, on each of the 1,000 samples, we implement the two EMCS procedures described in Section~\ref{sec:designs}, and compare their performances in terms of the criteria introduced in Section~\ref{sec:theory}.

One limitation of this approach is that the `true effect' we calculate is subject to sampling error. We therefore consider a second case, where we apply the insight of \cite{SmithTodd2005} that the \emph{control} sample from the NSW can be compared to the same non-experimental comparison group. The NSW control sample includes people who were selected in the same way as those actually treated, but who were randomised out of treatment. Now we know that the `true effect' is a precise zero, since the control sample did not actually receive treatment. Thus, we have an original dataset of 142 `treated' observations (who in reality received no treatment) and 7,467 comparison units. This comes from taking the `early random assignment' control sample from \cite{SmithTodd2005} and a version of the CPS-1 dataset trimmed to overlap with these controls. Again we draw samples by selecting 100 treated observations and 1,900 comparison observations from this population, with the true effect being precisely zero in each sample, and then perform EMCS on these samples.

Another possible worry might be that our example applies estimators that are suitable under unconfoundedness, \textit{i.e.}~when potential outcomes are independent of treatment assignment, conditional on observed covariates. \cite{SmithTodd2005} question this assumption in the context of the NSW--CPS data, and especially in the context of their `early random assignment' samples. To address this concern, we take a third approach. The basic idea is to construct a population similar to the NSW--CPS data where unconfoundedness holds by construction, and then draw samples from this. 
We begin with a trimmed version of the \cite{DehejiaWahba1999} dataset used in the first case. Next, we perform 4-nearest neighbour matching (with replacement) to impute the `missing' potential outcome for each observation. This is our new population, in which we have complete knowledge of both potential outcomes, as well as a propensity score for each observation estimated from the NSW--CPS data. 
We then draw random subsamples of 2,000 observations (covariates, potential outcomes, and propensity scores) from this artificially created population. 
For each observation we create a perturbed propensity score by adding a logistic error to the NSW--CPS estimated propensity score. We assign to treatment the individuals in the top quarter of the perturbed propensity score distribution (giving 500 treated and 1,500 nontreated in each sample). By construction treatment is therefore ensured to be independent of potential outcomes in this subsample. The overlap assumption is also satisfied, since the inclusion of a logistic error ensures that no individual is guaranteed to be treated. 
The true value of ATT in this sampling process is given by $\frac{1}{\Pr(D=1)} \cdot \E \left[ e(x) \cdot (Y(1)-Y(0)) \right]$, where $e(x)$ is the probability of being treated in the assignment rule based on the ranking of the perturbed propensity scores. By design we know $Y(1)$, $Y(0)$, and $\Pr(D=1)=.25$, and we can approximate $e(x)$ by the empirical frequencies in the simulations. 
Finally, we implement EMCS on the samples drawn in this way.

\subsection{Estimators}
\label{subsec:estimators}

In all our simulations we study the impact of the NSW programme on earnings in 1978. We consider seven nonexperimental estimators: linear regression, Oaxaca--Blinder, inverse probability weighting (IPW), doubly-robust regression, uniform kernel matching, nearest neighbour matching, and bias-adjusted nearest neighbour matching. For details see Appendix~\ref{sec:appendix_estimators}\@. In each case we focus on the average treatment effect on the treated (ATT), unless a given method does not allow for heterogeneity in effects (in which case we estimate the overall effect of treatment).
As noted above, all of these estimators are based on the assumption of unconfoundedness.

We use a single set of control variables in all our simulations. Following \cite{DehejiaWahba1999} and \cite{SmithTodd2005}, we control for age, age squared, age cubed, education, education squared, whether a high school dropout, whether married, whether black, whether Hispanic, earnings in months 13--24 prior to randomization, earnings in 1975, nonemployment in months 13--24 prior to randomization, nonemployment in 1975, and the interaction of education and earnings in months 13--24 prior to randomization.

We conduct all our simulations in Stata and use several user-written commands in our estimation procedures:~\texttt{nnmatch} (\citealt{ADHI2004}), \texttt{oaxaca} (\citealt{Jann2008}), and \texttt{psmatch2} (\citealt{LeuvenSianesi}).

\subsection{Procedures}
\label{subsec:procedures}

In Section~\ref{sec:designs} we noted that for the placebo design we require some choice of $\pi$ and $\lambda$, where $\lambda$ determines the degree of covariate overlap between the `placebo treated' and `placebo control' observations and $\pi$ determines the proportion of the `placebo treated'. We choose $\pi$ to ensure that the proportion of the `placebo treated' observations in each placebo EMCS replication is equal to the proportion of treated units in the sample.\footnote{It should be noted, however, that the way these datasets were constructed by \cite{LaLonde1986} results in samples that are best described as choice-based. More precisely, the treatment and control groups are heavily overrepresented relative to their population proportions. See \cite{SmithTodd2005} for a further discussion of this issue.} We also follow \cite{HLW} in choosing $\lambda=1$ as well as in using a logit model to estimate the propensity score.

The structured design requires more choices, in particular how we specify the joint probability distribution as the product of the marginal distribution for treatment status and some conditional distributions. As discussed in Section~\ref{sec:designs}, we begin each structured EMCS replication by generating a fixed number of treated and nontreated observations to match the numbers in the sample. We then order the covariates, regress each covariate on the preceding covariates (using logistic regression for binary covariates), and use this to define the conditional distribution for that covariate. In EMCS replications the covariates are then drawn in the same order, from the appropriate conditional distribution. Full details of the procedure are provided in Appendix~\ref{sec:appendix_strEMCSprocedure}.

%% file: results.tex
We now describe the results of our tests of the two EMCS procedures -- placebo and structured -- in the context of our real-world data. As described in Subsection~\ref{subsec:approach}, we perform three sets of tests. First, we apply the two procedures to the NSW treatment sample, combined with the CPS-1 comparison dataset. We find the performance of the procedures to be poor when it comes to finding the estimator with the lowest bias. When we study MSE (\textit{i.e.}, account also for variance), performance is better. This is because the rankings of estimators are mainly being driven by the variance, and both EMCS methods do well at replicating the variance components. However, given this, we also test a simple bootstrap procedure and find that it is more effective at picking the best estimator.
Then, we follow \cite{SmithTodd2005} in using the NSW controls as our `treated' sample instead: now the effect we intend to estimate must be zero for sure, removing worries that poor performance might be an artefact caused by sampling uncertainty around the true effect. We find that the previous results are maintained.
Finally, we use an adjusted version of the original data, constructed so that conditional independence necessarily holds, to allay concerns that poor performance is driven by a context in which unconfoundedness may not hold. Again we find that the EMCS procedures do not perform well on bias, and are better on MSE, although here the bootstrap does not clearly dominate.

\subsection{Testing EMCS in the NSW Data}
\label{subsec:EMCSinNSW}

Our first results using `real-world' data focus on the variant of the original NSW treatment sample constructed by \cite{DehejiaWahba1999}, combined with a trimmed version of the CPS-1 comparison dataset. 
We create 1,000 samples from the original dataset by sampling 100 treated and 1,900 nontreated observations from the 185 possible treated and 7,660 comparison units in the original data. 
We implement the two EMCS procedures 1,000 times on each of the 1,000 samples, giving a total of 1,000,000 replications for each EMCS procedure. 
In each replication we implement the seven estimators described earlier, and measure how well the two EMCS procedures help us assess the relative performance of the estimators. We might measure performance of an estimator in terms of absolute bias or MSE (which also takes into account its variance). Performance of EMCS (`EMCS-validity') is then measured by how well the EMCS procedure replicates these features of an estimator in the original samples. In Section~\ref{sec:theory}, we described two measures of EMCS performance suitable for when we have many estimators: 
\begin{enumerate}
 	\item the average regret, \textit{i.e.}~average difference in absolute bias/MSE between the estimator selected by EMCS and the estimator with the actual minimum absolute bias/MSE; and
 	\item the average Kendall's tau (Kendall's rank correlation coefficient), which measures the similarity between the ranking of estimators suggested by EMCS and the `true' ranking from the original samples.
\end{enumerate} 
For ease of interpretation, it is also useful to normalise the values of average regret. Our discussion below focuses on the average regret as a percentage of the minimum value of absolute bias/MSE\@. However, we also consider an alternative normalisation, where we divide the average regret for a given EMCS procedure by the average regret for random selection of estimators (which we discuss further below). Finally, we also consider an additional measure, which is straightforward to interpret, namely
\begin{enumerate}
	\item[3] the average correlation in absolute bias/MSE (rather than in the rankings, as given by Kendall's tau).
\end{enumerate}
In each case the comparison is between what the EMCS procedure suggests and the results from taking the `true effect' in the original data, and then calculating the absolute bias/MSE of each estimator across the 1,000 samples.

To provide a benchmark for the performance of EMCS, we also include results from two other procedures. In the first we simply apply nonparametric bootstrap to the same samples used for the EMCS procedures.\footnote{Precisely, we sample with replacement, and draw replication samples of the same size as the original sample. In consequence, our bootstrap samples are also of the same size as the structured samples. Both are larger than the placebo samples, which are of the size of the original \emph{comparison} subsample.} We can then compare estimators on absolute bias, variance, or MSE, and also see how the resulting rankings compare to those from the original samples. Our estimates of absolute bias and MSE are centred around the point estimates in each original sample. In the second we do not create any samples, but simply rank estimators randomly. This provides a `worst-case' benchmark: suppose a researcher knows nothing at all about performance and just picks an estimator blindly, how would they do? Here we cannot compute a result for the correlation, but can for average regret and Kendall's tau.
Table~\ref{tab:table1} shows the results from these simulations. Appendix~\ref{sec:appendix_tables} in \cite{AKS2019_SA} provides further details.

\input{table1}

The first result is that performance of both EMCS procedures in terms of bias is very poor. The average regret in terms of absolute bias, as a percentage of the absolute bias for the best estimator, is 3,067\% (3,766\%) for placebo (structured), \textit{i.e.}~an order of magnitude larger than the minimum value. 
It is worse than choosing completely randomly, which would be 1,184\% worse than the best estimator, and worse than the bootstrap, 1,000\%. 
Looking at the ranking across estimators, the average Kendall's tau is --.21 (--.37) for placebo (structured). So the rankings produced by EMCS are, on average, \emph{negatively} correlated with the ranking in the original samples. This is worse than random, which gives .00, and bootstrap, .02. The same pattern is seen in the average correlation coefficients for absolute bias, which are --.44 (--.51).

A researcher might be interested in knowing about performance of estimators in terms of MSE rather than only considering bias. Here EMCS performs much better. The average regret for placebo (structured) is now only 18\% (16\%), much better than random (142\%). Similarly, average Kendall's tau is now .60 and .64 for placebo and structured, respectively, much better than .00 for random. The lowest panel of Table~\ref{tab:table1} shows that this is driven by the much better performance in replicating the variances. Since the rankings here are mostly determined by the variance, being able to reproduce variances substantially improves the measures of performance relative to the metrics based on absolute bias.

However, looking at our other benchmark case -- the bootstrap -- we see that it outperforms both EMCS methods in terms of MSE\@. Average regret is lower at 7.9\%, and the average Kendall's tau is much higher at .83. Given that MSE performance for EMCS is driven by the variance components, this does not seem surprising. The bootstrap is a simpler procedure than the two EMCS methods, and its ability to help us understand the variability of estimators is well known. It therefore seems like a potentially valuable path which has fewer design choices than EMCS.

\subsection{Removing Sampling Error from the `True Effect'}
\label{subsec:RemovingSamplingError}

The previous subsection calculated the MSE for each estimator by comparing the value of the estimate in each sample to a `true effect' measured using the experiment. One concern might be that the estimate from the experiment is subject to sampling error, and this might somehow negatively affect our performance measures for EMCS\@. To test this, we now use as our `treated' observations the `early random assignment' NSW control sample from \cite{SmithTodd2005}. Since these individuals were selected for the programme in the same way as those actually treated, but were then randomised out, the actual treatment effect for them is precisely zero. We therefore repeat the exercise on these data, again implementing the two EMCS procedures 1,000 times on each of the 1,000 original samples. Table~\ref{tab:table2} documents the results. Appendix~\ref{sec:appendix_tables} in \cite{AKS2019_SA} provides further details.

\input{table2}

Our conclusions are similar to those in the previous subsection. In terms of absolute bias, the average regret is much lower than previously, at 30\% (42\%) for placebo (structured). However, this is mostly driven by a large increase in the minimum value of absolute bias, since it is much more difficult to recover the true effect of NSW in these data \citep{SmithTodd2005}. This can alternatively be seen from normalising values by the average regret for random selection of estimators. In the first simulation study, the average regret for placebo (structured) is 2.6 (3.2) times larger than for random; in the second, it is 1.6 (2.2) times larger. While these values continue to be smaller in the second simulation study, their overall magnitudes are similar in both cases. This also makes it clear that EMCS is still worse than choosing at random (average regret of 19\%) and bootstrap (10\%). As before, the average Kendall's tau is negative for placebo (structured) at --.27 \mbox{(--.47)}, which is worse than random (.00) and bootstrap (.32) as well. On MSE performance is better, with average regret of 23\% (32\%) and average Kendall's tau of .65 (.55). These are much better than random (94\% and .00), but worse than bootstrap (17\% and .81).

\subsection{Ensuring Unconfoundedness Holds}
\label{subsec:EnsuringUnconfoundednessHolds}

Another potential concern is whether the conditional independence assumption holds. Here we take the approach described in Subsection~\ref{subsec:approach} to generate 1,000 samples in which conditional independence holds by construction. Then, we implement the two EMCS procedures 500 times on each of these samples. Table~\ref{tab:table3} displays the results. Appendix~\ref{sec:appendix_tables} in \cite{AKS2019_SA} provides further details.

\input{table3}

The previous results are broadly maintained even after ensuring conditional independence. In terms of absolute bias the performance of both EMCS approaches is similar to random, though now slightly better than bootstrap. In terms of MSE both procedures perform better than random selection of estimators and also marginally better than bootstrap. Average regret in terms of MSE is worse than in the first case, though average Kendall's tau is a little higher, so it is also not obvious that contexts where conditional independence holds should necessarily see better performance of EMCS procedures.

%% file: table1.tex
\begin{table}[!t]
\begin{adjustwidth}{-1in}{-1in}
  \centering
  \caption{\bf EMCS-validity using different performance metrics} \label{tab:table1}
  \begin{threeparttable}
  	\begin{tabular}{lccccc}
  	  \toprule
  	  EMCS approach              & Placebo & Structured & Bootstrap & Random  \\
  	  \midrule
      \textbf{Absolute bias} (minimum = 16) & & & & \\
      Average regret (\% of minimum)
                    & 3,067    & 3,766    & 1,000  & 1,184   \\
      Average regret (\% of random)
                    & 259.0    & 318.1    &  84.5  & 100.0   \\
      Average Kendall's tau
                    & --.214   & --.372   &  .016  & 0  \\
      Average correlation  
                    & --.437   & --.505   &  .275  & ---        \\
                    \\
      \textbf{Mean squared error} (minimum = 512,322) & & & & \\
      Average regret (\% of minimum)
                    &  18.2    &  16.3    &  7.9   &  141.9  \\
      Average regret (\% of random)
                    & 12.8    & 11.5    & 5.6  & 100.0   \\
      Average Kendall's tau 
                    &  .599    &  .635    &  .828  & 0   \\
      Average correlation 
                    &  .647    &  .791    &  .809  & ---        \\
                    \\
      \textbf{Variance} (minimum = 454,278) & & & & \\
      Average regret (\% of minimum)
                    &  2.7     &  11.2    &  1.9   &  148.0  \\
      Average regret (\% of random)
                    & 1.8    & 7.6    & 1.3  & 100.0   \\
      Average Kendall's tau
                    &  .767    &  .812    &  .883  &  0      \\
      Average correlation 
                    &  .895    & .920     &  .862  & ---        \\
  	  \bottomrule
  	\end{tabular}
  	\begin{footnotesize}
  	\begin{tablenotes}[flushleft]
    \item \input{tablenotes}
  	\end{tablenotes}
  	\end{footnotesize}
  \end{threeparttable}
\end{adjustwidth}
\end{table}

%% file: tablenotes.tex
\textbf{Notes:} `EMCS approach' denotes the way in which the empirical Monte Carlo samples were generated. `Placebo' and `Structured' generate samples using the placebo and structured approaches described in Section~\ref{sec:designs}. `Bootstrap' generates nonparametric bootstrap samples by sampling with replacement the same number of observations as the original data. `Random' does not generate samples but instead randomly assigns \emph{rankings} to the estimators (hence statistics are only available for the performance metrics based on rankings).
The absolute bias, mean squared error, and variance are features of estimators. The `minimum' value for each feature is its lowest value among our estimators in the original data generating process (\textit{i.e.}~we have one value of each feature for each estimator in the `original samples' and we report the lowest of these values). See Appendix~\ref{sec:appendix_tables} in \cite{AKS2019_SA} for more details.
Four performance measures are used for each of these statistics.
`Average regret' measures the average increase in the statistic from choosing the estimator actually selected by the EMCS approach rather than the estimator with the minimum value of this statistic, as a percentage of \emph{(i)} that minimum value or \emph{(ii)} the average regret for random selection of estimators.
`Average Kendall's tau' measures the average correlation in the ranking of estimators provided by the EMCS approach relative to the ranking in the original samples.
`Average correlation' measures the average correlation in the actual values of the statistic (rather than the ranking) provided by the EMCS approach relative to the values in the original samples.
All averages are taken with respect to 1,000 original samples; for each sample, a separate simulation study was conducted.
The results for random selection of estimators are analytical; instead of actually generating random rankings, we report the known values of expected Kendall's tau (zero) and expected regret with random rankings. The latter value is equal to the average regret across estimators, with an equal probability of each estimator to be selected as `best'.

%% file: table2.tex
\begin{table}[!t]
\begin{adjustwidth}{-1in}{-1in}
  \centering
  \caption{\bf EMCS-validity ensuring no sampling error in the treatment effect} \label{tab:table2}
  \begin{threeparttable}
  	\begin{tabular}{lccccc}
  	  \toprule
  	  EMCS approach              & Placebo & Structured & Bootstrap & Random  \\
  	  \midrule
      \textbf{Absolute bias} (minimum = 954) & & & & \\
      Average regret (\% of minimum)
                    & 30.2    &  41.6   &  10.4   &  19.0   \\
      Average regret (\% of random)
                    & 158.9    & 219.1    & 54.9  & 100.0   \\
      Average Kendall's tau
                    & --.274  &  --.466 &  .320    &  0   \\
      Average correlation  
                    & --.418  &  --.822 &  .263   & ---        \\
                    \\
      \textbf{Mean squared error} (minimum = 1,222,627) & & & & \\
      Average regret (\% of minimum)
                    & 22.5    &  32.1   &  17.4   &  94.4   \\
      Average regret (\% of random)
                    & 23.9    & 34.0    & 18.4  & 100.0   \\
      Average Kendall's tau 
                    & .645    &  .549   &  .809   &  0 \\
      Average correlation 
                    & .843    &  .814   &  .746   & ---        \\
                \\
      \textbf{Variance} (minimum = 296,671) & & & & \\
      Average regret (\% of minimum)
                    & 1.2     &  5.0    &  9.0    &  262.6  \\
      Average regret (\% of random)
                    & .5    & 1.9    & 3.4  & 100.0   \\
      Average Kendall's tau
                    & .950     &  .665   &  .762   &  0      \\
      Average correlation 
                    & .959    &  .934   &  .833   & ---        \\
      \bottomrule
  	\end{tabular}
  	\begin{footnotesize}
  	\begin{tablenotes}[flushleft]
    \item \textbf{Notes:} See Table \ref{tab:table1}.
  	\end{tablenotes}
  	\end{footnotesize}
  \end{threeparttable}
\end{adjustwidth}
\end{table}

%% file: table3.tex
\begin{table}[!t]
\begin{adjustwidth}{-1in}{-1in}
  \centering
  \caption{\bf EMCS-validity ensuring unconfoundedness holds} \label{tab:table3}
  \begin{threeparttable}
  	\begin{tabular}{lccccc}
  	  \toprule
  	  EMCS approach              & Placebo & Structured & Bootstrap & Random  \\
  	  \midrule
      \textbf{Absolute bias} (minimum = 68) & & & & \\
      Average regret (\% of minimum)
                    &  593.5     &  670.1   &  972.1      &  522.0      \\
      Average regret (\% of random)
                    & 113.7    & 128.4    & 186.2  & 100.0   \\
      Average Kendall's tau
                    &  --.003  &  .057  &  --.300     &  0   \\
      Average correlation  
                    &  --.048  &  .217  &  --.263   & ---          \\
                    \\
      \textbf{Mean squared error} (minimum = 340,300) & & & & \\
      Average regret (\% of minimum)
                    &  260.1     &  89.7    &  276.1      &  682.9      \\
      Average regret (\% of random)
                    & 38.1    & 13.1    & 40.4  & 100.0   \\
      Average Kendall's tau 
                    &  .737    &  .729  &  .632     &  0   \\
      Average correlation 
                    &  .943    &  .790  &  .778     & ---          \\
                \\
      \textbf{Variance} (minimum = 137,574) & & & & \\
      Average regret (\% of minimum)
                    &  0       &  .3  &   0       &  1,631   \\
      Average regret (\% of random)
                    & 0    & 0    & 0  & 100.0   \\
      Average Kendall's tau
                    &  .768    &  .727  &  .752     &  0        \\
      Average correlation 
                    &  .951    &  .820  &  .820     & ---          \\
      \bottomrule
  	\end{tabular}
  	\begin{footnotesize}
  	\begin{tablenotes}[flushleft]
    \item \textbf{Notes:} See Table \ref{tab:table1}.
  	\end{tablenotes}
  	\end{footnotesize}
  \end{threeparttable}
\end{adjustwidth}
\end{table}

%% file: discussion.tex
Advances in econometrics have left the empirical researcher blessed with a wealth of possible treatment effect estimators from which to choose. They have not yet provided clear guidance on which of these estimators should be preferred in which context. In this paper we studied two proposals which suggest an approach to choosing an appropriate estimator for a given context. The first approach (placebo) suggests a way to introduce placebo treatments to some control observations in a dataset, and studies how well estimators can pick up the true zero effect. The second approach (structured) creates data from a known DGP whose parameters are estimated from features of the original data, and studies how well estimators can pick up the implied true effect in the DGP.

We showed theoretically that both approaches can only be guaranteed to work under rather restrictive conditions. Specifically, when they can correctly reproduce both the biases and the ordering of the variances of estimators. We showed simple practical cases where one or other of these might fail, and gave an example of the consequences based on simulations from an artificial DGP\@. To provide a real-world example, we also implemented the EMCS procedures in the NSW--CPS data, where we know the `true effect' of the programme. This allowed us to compute actual performance of the estimators in samples from the original data, and compare this to what EMCS would suggest if applied to these samples. We showed that in this example EMCS performs badly on ordering estimators in terms of absolute bias, and the estimator it suggests is often many times worse than the best (or even than selecting randomly). In this example both EMCS procedures perform much better in terms of MSE because reproducing the variance term turns out to drive the MSE in these data. But, this leads the methods to be no better (and sometimes substantially worse) than a simple bootstrap procedure.

These results are unfortunate, but nevertheless important. There remains no silver bullet that can assist empirical researchers with the `right' or `best' estimator for a particular context. In the absence of a clear choice driven by research design, the best advice at this stage is likely to be implementing a number of estimators, and then considering the range of estimates provided, as \cite{BussoDiNardoMcCrary_PSMandReweighting} also suggest.

One possible future alternative, recently proposed, is \textit{synth-validation} \citep{SynthVal}. This approach is related to cross-validation and is based on estimating `the estimation error of causal inference methods applied to a given dataset'. The authors provide simulations which suggest that this `lowers the expected estimation error relative to consistently using any single method'. Further work is needed to test how general this approach is, and whether it can reliably guide researchers in selecting estimators.

%% file: appendix_theory.tex
\textbf{Derivations of (\ref{eq:SEB1}) and (\ref{eq:sigma2})}: A general expression of SEB for ATT in the absence of knowledge of the propensity score is given by
\begin{equation}
SEB_{ATT} = \frac{1}{\Pr(D=1)} E \left[ Var(Y(1)|X) + \frac{e(X)}{1-e(X)} Var(Y(0)|X) + (\tau(X) - \theta^0 )^2 | D=1 \right]. \notag
\end{equation} 
Plugging the current specifications for $Var(Y(1)|X)$ and $Var(Y(0)|X)$ and noting $\tau(X) = \theta^0$ for all $X$, the expression of (\ref{eq:SEB1}) follows.

By the partialling out argument of the least squares regression and the linear probability specification of the propensity score, the asymptotic variance of $\sqrt{n}(\hat{\theta}_2 - \theta^0)$ can be written as
\begin{align}
Avar(\sqrt{n}(\hat{\theta}_2 - \theta^0)) & = \frac{E(\epsilon^2(D-e(X))^2)}{[E((D - e(X))^2)]^2} \notag \\
&=\frac{E\left[ Var(Y(1)|X) (1-e(X))^2 e(X) + Var(Y(0)|X) e(X)^2 (1-e(X)) \right] }{ [E(e(X)(1-e(X))]^2} \notag \\
& = \frac{\sigma^2}{\Pr(D=1)} \cdot \frac{E \left[ c  (1-e(X))^2 + e(X)(1-e(X)) | D=1 \right] }{[E(1-e(X) | D=1)]^2}, \notag
\end{align}  
where the third line follows from Bayes rule applied to each denominator and numerator. $\square$

\bigskip

\noindent \textbf{Proof of (\ref{eq:condition_c})}: Rewrite (\ref{eq:SEB1}) and (\ref{eq:sigma2}) as
\begin{align}
\sigma_1^2 & = \frac{\sigma^2}{\Pr(D=1)} \left[ c -1 + E\left( \frac{1}{1-e(X)} |  D=1 \right)\right], \label{SEB3} \\
\sigma_2^2 & = \frac{\sigma^2}{\Pr(D=1)} \left[ (c-1) \cdot \frac{E((1-e(X))^2 | D=1)}{[E(1-e(X)|D=1)]^2} + \frac{1}{E(1-e(X)|D=1)}\right]. \label{sigma2_2}
\end{align}
Hence, we obtain 
\begin{equation}
\sigma_2^2 - \sigma_1^2 = \frac{\sigma^2}{\Pr(D=1)}\left[ (c-1) \Delta_2 - \Delta_1 \right],
\end{equation}
$\Delta_1 \geq 0$ by Jensen's inequality, and $\Delta_2 \geq 0$. Hence, the condition for $\sigma_2^2 > \sigma_1^2$ follows as in (\ref{eq:condition_c}).
$\square$

%% file: appendix_estimators.tex
We use seven estimators in our empirical application.

\begin{enumerate}
	\item Linear regression (OLS).
	\item Oaxaca--Blinder -- we follow \citet{Kline2011} in using the Oaxaca--Blinder decomposition to estimate the ATT\@.
	\item Inverse probability weighting (IPW) -- we first estimate the propensity score using a logit model, and then use inverse weighting with normalised weights to estimate the ATT\@.
	\item Doubly-robust regression -- as in \cite{Wooldridge2007} and \cite{SW2018}, we use the inverse-probability-weighted regression-adjustment (IPWRA) estimator. This is effectively a combination of the two estimators above, IPW and Oaxaca--Blinder. It satisfies the double robustness property.
	\item Uniform kernel matching -- we first estimate the propensity score using a logit model, and then match on propensity scores using a uniform kernel. We select the bandwidth on the basis of leave-one-out cross-validation (as in \citealt{BussoDiNardoMcCrary_FiniteSample} and \citealt{HLW}), using a search grid $.005 \times 1.25^{g-1}$ for $g=1,2,\ldots,15$. The computational time of doing this for each replication is prohibitive. Consequently we calculate this once for each original sample, and use the recovered optimal bandwidth in all EMCS replications for that sample.
	\item Nearest neighbour matching -- nearest neighbour matching on propensity scores, which are first estimated from a logit regression, with matching on the single nearest neighbour. We match with replacement; if there are ties, all of the tied observations are used.
	\item Bias-adjusted nearest neighbour matching -- as above, but correcting bias as in \cite{AbadieImbens2011}, since nearest neighbour matching is not $\sqrt{n}$-consistent. 
\end{enumerate}

%% file: appendix_strEMCSprocedure.tex
Here we detail precisely the procedure followed to implement the structured EMCS in our empirical application. As noted previously, we begin each structured EMCS replication by generating a fixed number of treated and nontreated observations to match the number in the sample. We then draw an employment status pair of \texttt{u74} and \texttt{u75} (nonemployment in months 13--24 prior to randomization and nonemployment in 1975), conditional on treatment status, to match the observed conditional joint probability. For individuals who are employed in only one period, an income is drawn from a log normal distribution conditional on treatment and employment statuses, with mean and variance calibrated to the respective conditional moments in the data. Where individuals are employed in both periods a joint log normal distribution is used, again conditioning on treatment status. In all cases, whenever the income draw in a particular year lies outside the relevant support observed in the data, conditional on treatment status, the observation is replaced with the limit point of the empirical support, as also suggested by \citet{BussoDiNardoMcCrary_PSMandReweighting}.

We model the joint distribution of the remaining control variables as a particular tree-structured conditional probability distribution, so that we can better fit the correlation structure in the data. The process for generating these covariates is as follows:
\begin{enumerate}
\item The covariates are ordered: treatment status, employment statuses, income in each period, whether a high school dropout (\texttt{nodegree}), education (\texttt{educ}), age, whether married, whether black, and whether Hispanic. This ordering is arbitrary, and a similar correlation structure would be generated if the ordering were changed.
\item Using the sample on which the EMCS is being performed, each covariate from \texttt{nodegree} onward is regressed on all the covariates listed before it (we use the logit model for binary variables).\footnote{One exception is \texttt{educ} which is regressed on the prior listed covariates conditional on \texttt{nodegree}. Clearly, it is not possible for a high school dropout to have twelve years of schooling or more; it is also not possible for a non-dropout to have less than twelve years of schooling.} These regressions are not to be interpreted causally; they simply give the conditional mean of each variable given all preceding covariates.
\item In the simulated dataset, covariates are drawn sequentially in the same order. For binary covariates a temporary value is drawn from a $\mathcal{U}[0,1]$ distribution. Then the covariate is equal to one if the temporary value is less than the conditional probability for that observation. The conditional probability is found using the values of the existing generated covariates and the estimated coefficients from step 2. Age and education are drawn from a normal distribution whose mean depends on the other covariates and whose variance is equal to that of the residuals from the relevant model. Again, we replace extreme values with the limit of the support, conditional on treatment status (for education, also conditional on dropout status).
\end{enumerate}
The outcome studied is earnings in 1978, \texttt{re78}. The simulated outcome, $Y_{i}$ for individual $i$, is then generated in two steps. In the first step, we generate a conditional mean using the parameters of a flexible linear model fitted to the sampled data. Precisely, we estimate $(\boldsymbol{\delta}_{0},\boldsymbol{\delta}_{1})$ from the following linear model:
\begin{equation}
\e ( Y \vert D, \mathbf{X} ) = (1-D)\mathbf{X}\boldsymbol{\delta}_{0} + D\mathbf{X}\boldsymbol{\delta}_{1}.
\label{eq:condmean}
\end{equation}
The predicted conditional mean in the replication is then calculated using the estimated coefficients $(\boldsymbol{\hat{\delta}}_{0},\boldsymbol{\hat{\delta}}_{1})$ from above, and the simulated treatment status and covariates, $D_{i}$ and $\mathbf{X}_{i}$. In the second step, the simulated outcome, $Y_{i}$, is determined as a draw from a normal distribution with the estimated conditional mean $(1-D_{i})\mathbf{X}_{i}\boldsymbol{\hat{\delta}}_{0} + D_{i}\mathbf{X}_{i}\boldsymbol{\hat{\delta}}_{1}$ and the variance that is fitted to that of the residuals from the model in equation (\ref{eq:condmean}), conditional on treatment status. Once again, we replace extreme values of \texttt{re78} with the limit point of the support, also conditional on treatment status. `True effects' in each replication, $\tilde{\theta}^0$, are calculated using the conditional means for both treatment statuses, and the difference in conditional means, \textit{i.e.}~the individual-level treatment effect, is averaged over the subsample of treated units. Thus, we implicitly focus on the sample average treatment effect on the treated (SATT), not on the population average treatment effect on the treated (ATT)\@. Both of these measures can be used as the benchmark effect in simulations and we have no particular preference for either.